# High quality, lightweight and adaptable TTS using LPCNet

*Zvi Kons, Slava Shechtman, Alex Sorin, Carmel Rabinovitz, Ron Hoory*

IBM research

{zvi , slava , sorin , hoory }@il.ibm.com, carmel.rabinovitz@ibm.com

## Abstract

We present a lightweight adaptable neural TTS system with high quality output. The system is composed of three separate neural network blocks: prosody prediction, acoustic feature prediction and Linear Prediction Coding Net as a neural vocoder. This system can synthesize speech with close to natural quality while running 3 times faster than real-time on a standard CPU.

The modular setup of the system allows for simple adaptation to new voices with a small amount of data.

We first demonstrate the ability of the system to produce high quality speech when trained on large, high quality datasets. Following that, we demonstrate its adaptability by mimicking unseen voices using 5 to 20 minutes long datasets with lower recording quality. Large scale Mean Opinion Score quality and similarity tests are presented, showing that the system can adapt to unseen voices with quality gap of 0.12 and similarity gap of 3% compared to natural speech for male voices and quality gap of 0.35 and similarity of gap of 9 % for female voices.

## 1. Introduction

In recent years we are experiencing a dramatic improvement of the synthesized speech quality in TTS systems, with the introduction of systems that are based on neural networks (NN). A major improvement in quality was achieved by using attention based models such as Tacotron [1] and by replacing vocoders with a NN based waveform generators such as WaveNet [2].

A useful feature of systems with trainable models is the ability to adapt the TTS to an unseen voice using a small amount of training data (from a few seconds to an hour of speech). This is usually done by training the system on a large number of speakers, and providing a speaker embedding vector as one of the system's inputs. Using this approach allows later retraining of only a subsets of the model parameters or prediction of the speaker embedding vector [3], [4], [5].

The drawback of this approach is that the resulting systems use large NN models. Furthermore, a multi-speaker model usually needs much more trainable parameters than a single speaker model. This may lead to a computationally heavy and slow synthesis process even on a strong GPU. Such requirements pose a severe problem for practical TTS system that require very low latency for a dialog with a human.

In our previous paper [6] we introduced a NN based TTS system with two trainable modules for prosody prediction and acoustic features prediction. This system used the WORLD vocoder [7]. We demonstrated that this TTS allows simple adaptation to new voices. This was carried out by retraining NN models that had already been trained using a large high-quality voice, on a small amount of data from the new voice. Although the quality of this system was better in many cases than similar concatenative TTS, it was still limited by the quality of the WORLD vocoder.

Recently, an efficient neural vocoder called LPCNet was introduced [8]. The LPCNet inference runs faster than real-time on a single CPU while producing a high quality speech output. LPCNet uses cepstrum representing spectral envelopes, pitch and pitch correlation as input features. This makes it a simple alternative to other vocoders, e.g. WORLD, which work with similar features.

In this paper we show that we can get a considerable quality improvement by modifying a TTS system that produced the WORLD vocoder parameters [6] to predict parameters for LPCNet [8]. As in the previous work [6], we conduct multiple adaptation experiments, applied on multiple VCTK voices [9] and show that the new system has much better quality and similarity to the target voices but can still run much faster than real-time in a single-CPU mode.

## 2. System architecture

An overview of our new TTS system is presented in Figure 1. The system is a cascade of a rule-based front-end, a NN based prosody generator, a NN synthesizer and an LPCNet decoder.

We adopted the front-end block which is used in the IBM Watson TTS engine and is described in detail in [10]. The front-end performs a grapheme-to-phoneme conversion, represents each word with a set of positional and categorical linguistic features and associates the features with the phonemes contained within the word.

The prosody generator is described in section 2.1. It emits a sequence of sub-phoneme elements, including duration, pitch and intensity values. Each sub-phoneme element represents either a heading, a middle or a trailing part of a phoneme.

The synthesizer is described in section 2.2. It represents each sub-phoneme element by several consecutive frames according to the element's duration and generates an acoustic feature vector for each frame.

Finally, an LPCNet block (section 2.3) is used to convert the stream of the acoustic feature vectors to a speech signal.

The prosody generator, synthesizer and LPCNet blocks use neural-net models for generating their output. Each block has its own model which is trained independently for each voice. Hence, the system is modular and provides easy control, flexibility and adaptability at the component level.

For each voice, the training and adaptation phases include the following data pre-processing steps:

1. A grapheme-to-phoneme conversion using the front-end block.
2. Forced alignment of audio at the sub-phoneme level using proprietary acoustic modeling and speech recognition tools.
3. Extraction of textual features for prosody modeling using the front-end block.
4. Pitch detection for prosody modeling using a proprietary tool.
5. Cepstra and residual extraction using the LPCNet feature extraction tool.

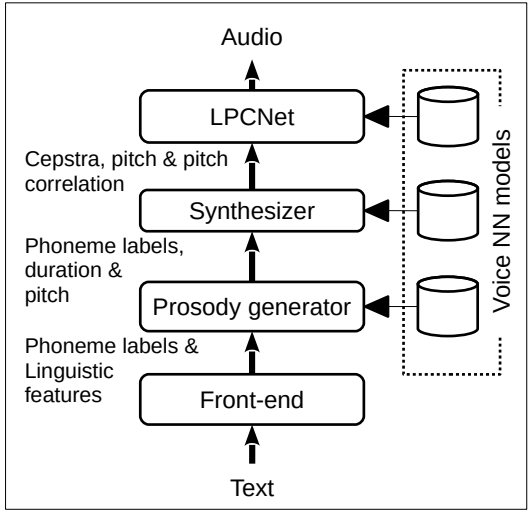

Figure 1: *TTS system architecture*

### 2.1. Prosody generator

In the current work, the prosody generation and adaptation network follows the one presented in our previous work [6], where one can refer to for more details. It generates a 4-dimensional prosody vector per TTS unit, comprising the unit's log-duration, initial log-pitch, final log-pitch and log-energy. The TTS units correspond to roughly 1/3 of a phone and result from forced-alignments with 3-state hidden Markov models. The input features, derived from the TTS Front End, are comprised of 1-hot coded categorical features and standard positional features [10].

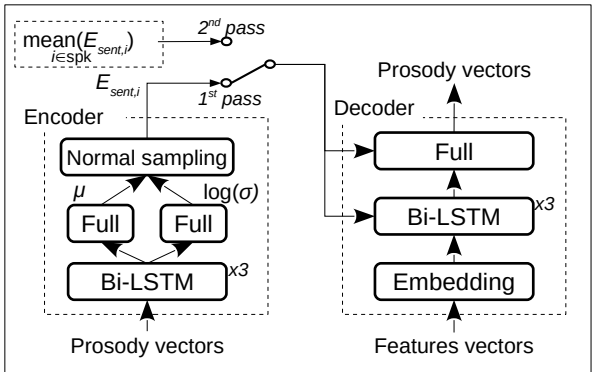

Figure 2: *Prosody generator training and retraining*

In this architecture the prosody adaptation to unseen speaker is based on a Variational Auto Encoder (VAE) utterance prosody embedding, averaged over all the speaker utterances [6], as presented on figure 2. In the current work we used multi-speaker baseline models for prosody adaptation to unseen voices, as it resulted in better quality than the single speaker models.

### 2.2. Synthesizer

The synthesis process begins by resampling the phonetic data and pitch to 10msec frames based on their duration predicted by the prosody generator. The sub-phoneme labels are represented by 32 element vectors, using a trainable embedding table.

Time dependencies and local context are extracted by convolution layers. The convolution is performed over time on the phonetic vector and the pitch curves independently with a window size of 0.32sec (forward and backward in time).

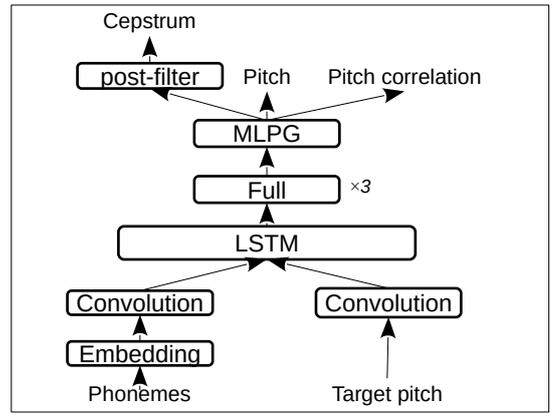

Figure 3: *Synthesizer network*

A longer time dependent context is extracted by an LSTM layer that merges the phonetic and pitch context. Following this are 3 fully connected layers with RELU non-linearity.

From the top layer we generate by linear transformations the speech parameters that the LPCNet requires as input: 18 cepstral vector, pitch and a pitch correlation parameters with first and second derivatives for all (total of 60 parameters).

The final parameters which we use as input for the LPCNet are found by solving the Maximum Likelihood Parameter Generation (MLPG) equations [12]. We also apply a formant enhancement filter on the cepstral coefficients $C_k$, $k=1...N$ to compensate for the NN averaging and to improve the speech quality similar to [13]. The enhancement starts by multiplication of the high-order coefficients:

$$C'_k = \begin{cases} C_k & k < K \\ \alpha C_k & k \geq K \end{cases} \quad (1)$$

We choose $\alpha=1.4$ and $K=2$. This can cause changes to the energy of the output, so we have to normalize it. Let $E[C]$ be the energy of the signal derived from the coefficients $C$. To calculate $E[C]$ we convert back from cepstrum to power spectrum and apply the inverse pre-emphasis filter. The energy is now the integration of this power spectrum.

Finally, to compensate for the energy change, we apply:

$$C'_0 = C_0 + \sqrt{N} \log_{10}\left(\frac{E[C]}{E[C']}\right) \quad (2)$$

The architecture of the synthesizer is shown in figure 3. The size of the layers is: phonetic embedding: 32, phonetic convolution: 128, pitch convolution: 32, LSTM: 512 and full: 512.

The network is trained using an aligned corpus where the inputs are the frame based phonetic labels and pitch values, while the outputs are the corresponding LPCNet parameters. We use MSE loss function on all output parameters.

We first train two single-speaker models form large male and female datasets (see section 3.1). Those are used as the base models for the adaptation experiments (section 3.2).

To adapt the model to a smaller unseen voice, we first initialize the training with the weights of the base model of the same gender. Then, the model is trained on a small target voice. A held-out validation set is used as a stop criterion for the adaptation to avoid over-fitting.

### 2.3. LPCNet decoder

The LPCNet decoder [8] is a WaveRNN [14] variant that uses a NN model to generate speech samples from equidistant-in-time input of cepstrum, pitch and pitch correlation parameters. Unlike other waveform generative models, such as WaveNet and WaveRNN, the LPCNet uses its NN to predict the LPC residual (the vocal source signal) and then apply to it an LPC filter calculated from the cepstrum.

This has the advantages of better control over the output of the spectral shape since it depends directly on the LPC filter shape. The model is also more robust to the predicted residual errors since any high frequency noise is also shaped by the LPC filter.

In this work we used the code published by the Mozilla team on Github[1] with some adjustments:

1. We replaced the pitch and pitch correlation values with values that were produced by our tools in order to maintain data consistency over all blocks.
2. We removed any data augmentation.
3. We added validation score over held-out data to the training procedure. This score was used to select the best model and served as a training stop criteria.

The LPCNet model was reported to perform well in speaker independent setting, when trained on multi-speaker datasets [8], however, we experimentally found that its performance further improves when retraining the initial multi-speaker same-gender model with the target voice specific data. The validation score (evaluated on 10% of held out validation data) was used to avoid over-fitting.

## 3. Experiments

Speech samples from the following experiments are available online at http://ibm.biz/IS2019TTS

### 3.1. High quality voices

For the first experiment we built a male and a female high-quality TTS systems. We used two proprietary datasets that were originally created for building a product level concatenative TTS system. The male dataset contains 13 hours of speech and the female dataset contains 22 hours of speech. Both were produced by native US English speakers and recorded in a professional studio. The audio was recorded sentence by sentence.

For each of those voices we built the following single speaker TTS systems:

1. A WORLD based system at 22KHz as described in [6].
2. An LPCNet based system at 16KHz as described in the previous section.
3. Tacotron2 based TTS with WaveNet decoder at 22Khz [1]

We used each one of these systems to synthesize a set of 40 held-out sentences and compared them by a MOS test to the original recordings. Because of the differences in sample rates, all of the samples were down-sampled to 16Khz and normalized to the same energy. The tests were performed using the Amazon Mechanical Turk (AMT) platform with 50-80 anonymous and untrained subjects participating in several evaluation sessions, constructed so that each sentence is evaluated by 30 distinct subjects. The quality MOS deployed a 5 points scale. The score for each system was calculated as the average over all its sentences. Table 1 shows the results of those tests. For the female voice the statistical significance of the difference between the LPCNet and the Tacotron systems is small (i.e. high $p$-value).

Table 1: *MOS quality for male and female voices*

| System | Male | Female |
|---|---|---|
| WORLD | 3.11 ± 0.06 | 2.99 ± 0.06 |
| LPCNet | 3.82 ± 0.05 | 3.75 ± 0.05 |
| Tacotron | 3.95 ± 0.05 | 3.86 ± 0.05 |
| Original | 4.32 ± 0.05 | 4.11 ± 0.05 |

We can see from these results that the LPCNet model has a huge impact on the quality compared to the WORLD system. We can also see that even though the LPCNet system has much lower complexity than the Tacotron2 like system, it gets close to it in quality.

One should note relatively low MOS scores for the original natural samples, which can be explained by the assumption that the listeners subjectively judged speaker pleasantness together with the speech quality and naturalness. Hence, the evaluation should be based on relative comparison between the different systems and the original samples.

### 3.2. Voice adaptation

In this experiment we selected 4 male and 4 female US English speakers from the VCTK [9] corpus. We created 3 datasets out of each voice: the first contains the entire available data (19 – 24 minutes of audio with average of 22 minutes), and the others two contain a random subset of the audio with total duration of 5 and 10 minutes.

From each one of these datasets we created a single voice TTS system. The networks for the acoustic features and the LPCNet where adapted from the corresponding, same gender networks that where trained in section 3.1. The prosody network was adapted from a multi-speaker baseline model (that was originally trained on high-quality voices and VCTK voices).

In addition, we also built a WORLD based TTS for each voice by adapting the WORLD based acoustic feature networks from the corresponding same gender networks of section 3.1 using the full voice data.

---

[1]https://github.com/mozilla/LPCNet

From each one of these systems we synthesized 40 sentences using text that was excluded from all the datasets. We evaluated each system's quality with MOS tests as in section 3.1. For reference, the tests also included 40 samples from the original VCTK datasets. We noticed that the original samples usually do not contain full sentences but rather short phrases. This factor, combined with the fact that VCTK comprises unprofessional speakers with varying voice pleasantness led, most probably, to the relatively low scores the original recordings received. The results of this test are summarized in Table 2. The statistical significance of the difference between the 5m and 10m LPCNet results is small.

Table 2: *MOS quality for adapted voices*

| System | Male | Female |
|---|---|---|
| WORLD 20m | 2.95 ± 0.09 | 2.51 ± 0.07 |
| LPCNet 5m | 3.88 ± 0.06 | 3.52 ± 0.06 |
| LPCNet 10m | 3.96 ± 0.06 | 3.64 ± 0.06 |
| LPCNet 20m | 4.02 ± 0.06 | 3.67 ± 0.06 |
| Original | 4.14 ± 0.06 | 4.02 ± 0.06 |

To measure the similarity of the synthesized voices to the original voices we performed additional subjective listening tests on AMT. In these tests a subject is presented with a pair of samples that convey different text messages. The subject is asked to rate their voice similarity, using a 4-point scale adopted from the Voice Conversion Challenge (VCC) [15] [16], and utilized in our previous experiments [6]. We performed two tests: one with only male voices and the second with only female voices. For reference, in each test we also checked the similarity of pairs of natural speech samples from the same speaker and of pairs of natural speech from different speakers of the same gender. The results are summarized in table 3. For each system we show the average score (on the scale 1-4) and the percentage of votes, which indicated that the two presented samples were from the same speaker (option 3 or 4). The statistical significance is small for the differences between all the male LPCNet systems and also for the difference between the female LPCNet 5m and 20m systems.

Table 3: *similarity of adapted systems*

| | Male | | Female | |
|---|---|---|---|---|
| System | Avg. Score | Same speaker | Avg. Score | Same speaker |
| Different | 2.08 ± 0.04 | 38% | 1.71 ± 0.03 | 21% |
| WORLD 20m | 2.94 ± 0.03 | 74% | 2.63 ± 0.04 | 58% |
| LPCNet 5m | 3.27 ± 0.03 | 84% | 3.21 ± 0.03 | 80% |
| LPCNet 10m | 3.29 ± 0.03 | 84% | 3.28 ± 0.03 | 81% |
| LPCNet 20m | 3.32 ± 0.03 | 85% | 3.15 ± 0.03 | 77% |
| Same | 3.41 ± 0.03 | 88% | 3.42 ± 0.03 | 86% |

We can compare these results to those of the VCC 2018 Hub task [15]. Although the task setup and the listening tests conditions are a bit different we can see that our system MOS and similarity score are comparable to those of the best VCC system (N10 with quality of 4.06 and Similarity of 85% where the corresponding scores for the original speech are 4.67 and 95%).

Figure 4 shows the results for each voice. To compensate for the variability between the voices, we have normalize the MOS scores for each voice in the range from 1 to the score of natural samples of this voice. The similarity scores for each voice, were normalized to the range between the scores of natural samples from different and same speakers. To clarify, the normalization ranges are different for each voice.

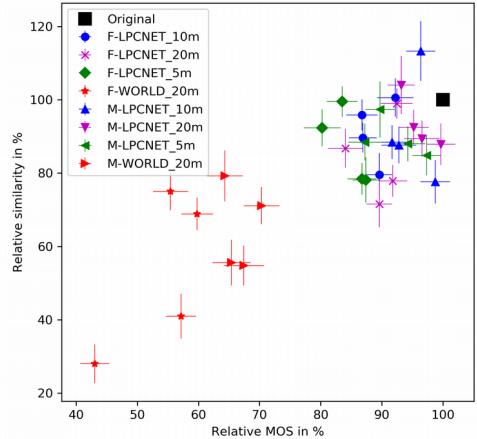

Figure 4: *relative MOS vs. relative similarity*

### 3.3. Performance

The slowest block of this TTS is the LPCNet. We found that it runs about 4 times faster than real-time on a 2.8GHz i7 CPU (no GPU was used). When adding the rest of the blocks we found that we can synthesize about 3 time faster than real-time on a CPU.

## 4. Conclusions

We have presented in this article a new TTS system that addresses the challenging goals of producing high quality speech while operating at faster than real-time rate without an expensive GPU support. The system is built around three NN models for generating the prosody, acoustic features and the final speech signal.

We tested this system using two proprietary TTS voice datasets and demonstrated that our system produces high quality speech that is comparable to larger and much slower Tacotron2 + Wavenet systems.

The task of creating a high-quality TTS system out of a smaller set of audio data is even more challenging. We have shown that our system can perform well even with datasets as small as 5-20 minutes of audio. We demonstrated that when we reduce the size of the training data, there is some graceful degradation to the quality, but we are still able to maintain good similarity to the original speaker.

For future work, we plan to allow voice modifications by adding control over voice parameters such as pitch, breathiness and vocal tract.